# DC superconducting quantum interference devices fabricated using bicrystal grain boundary junctions in Co-doped BaFe$_2$As$_2$ epitaxial films

Takayoshi Katase[1], Yoshihiro Ishimaru[2], Akira Tsukamoto[2], Hidenori Hiramatsu[3], Toshio Kamiya[1], Keiichi Tanabe[2] and Hideo Hosono[1,3*]

[1] Materials and Structures Laboratory, Mailbox R3-1, Tokyo Institute of Technology, 4259 Nagatsuta-cho, Midori-ku, Yokohama 226-8503, Japan

[2] Superconductivity Research Laboratory, International Superconductivity Technology Center, 10-13 Shinonome 1-chome, Koto-ku, Tokyo 135-0062, Japan

[3] Frontier Research Center, S2-6F East, Mailbox S2-13, Tokyo Institute of Technology, 4259 Nagatsuta-cho, Midori-ku, Yokohama 226-8503, Japan

(*) E-mail: hosono@msl.titech.ac.jp

**Abstract**

DC superconducting quantum interference devices (dc-SQUIDs) were fabricated in Co-doped $BaFe_2As_2$ epitaxial films on (La, Sr)(Al, Ta)$O_3$ bicrystal substrates with 30° misorientation angles. The 18 × 8 $\mu m^2$ SQUID loop with an estimated inductance of 13 pH contained two 3 μm wide grain boundary junctions. The voltage-flux characteristics clearly exhibited periodic modulations with $\Delta V$ = 1.4 μV at 14 K, while the intrinsic flux noise of dc-SQUIDs was 7.8 × $10^{-5}$ $\Phi_0$/Hz$^{1/2}$ above 20 Hz. The rather high flux noise is mainly attributed to the small voltage modulation depth which results from the superconductor-normal metal-superconductor junction nature of the bicrystal grain boundary.

## 1. Introduction

Since the discovery of Fe-pnictide superconductors [1, 2], intensive research on epitaxial films has been conducted to develop superconductor device applications as well as to examine their intrinsic properties. After our group demonstrated epitaxial films of LaFeAsO [3] and superconducting epitaxial films of cobalt-doped $SrFe_2As_2$ [4], several groups reported epitaxial film growth of Fe-pnictide superconductors with high critical temperatures ($T_c$) ≥ 20 K, including fluorine-doped $Ln$FeAsO ($Ln$ = La, Nd) [5-10], cobalt-doped $AE$Fe$_2$As$_2$ ($AE$ = Sr, Ba) [11-16], potassium-doped $BaFe_2As_2$ [17], and $FeSe_{1-x}Te_x$ [18].

Currently high quality films possessing a high critical current density ($J_c$) > 1 MA/cm$^2$ and high crystallinity have been realized only in cobalt-doped $BaFe_2As_2$ ($BaFe_2As_2$:Co) epitaxial films [15, 19]. Lee et al. have reported that bicrystal grain boundaries with large misorientation angles drastically suppress $J_c$ in $BaFe_2As_2$:Co epitaxial films grown on $SrTiO_3$ bicrystal substrates [13]. Recently, we succeeded in fabricating Josephson junctions in high-quality $BaFe_2As_2$:Co epitaxial films grown on (La, Sr)(Al, Ta)$O_3$ (LSAT) bicrystal substrates with 30° misorientation angles [19]; however, developing superconducting devices remains a challenge. Hence, examining the potential of Fe-pnictide superconductors is important.

Superconducting quantum interference devices (SQUIDs) are significant applications of Josephson junctions because they are useful as both highly sensitive magnetic sensors and tools used to evaluate film quality, to probe junction barrier properties, and to determine the symmetry of the superconducting order parameter. Chen and co-workers have observed the half-flux quantum effect due to the sign change of the order parameter in a fluorine-doped NdFeAsO polycrystalline bulk where the grain boundary Josephson junctions are connected by an external Nb wire [20]. However, the device characteristics could not be controlled because their SQUID loops contained naturally formed grain boundaries.

In this letter, we demonstrate the operation of dc-SQUIDs using a superconducting loop containing two Josephson junctions defined by bicrystal grain boundaries in $BaFe_2As_2$:Co epitaxial films.

## 2. Experimental details

Pulsed laser deposition (PLD) was used to fabricate 310 nm-thick cobalt-doped $BaFe_2As_2$ ($BaFe_2As_2$:Co) epitaxial films on [001]-tilt LSAT bicrystal substrates with 30° misorientation angles. The second harmonic of a Nd:YAG laser (wavelength: 532 nm) with a repetition rate of 10 Hz was used as the excitation source, and the PLD target was a stoichiometric $BaFe_{1.84}Co_{0.16}As_2$ polycrystalline disk [21]. The films were grown at the optimum growth temperature of ~850 °C [19]. The base pressure of our PLD chamber was $(1 - 2) \times 10^{-6}$ Pa, and film deposition was carried out in a vacuum at ~$10^{-5}$ Pa. The obtained $BaFe_2As_2$:Co films were composed of an almost pure $BaFe_2As_2$:Co phase [19]. The films have an epitaxial relationship with the LSAT bicrystal substrate; i.e. the $BaFe_2As_2$:Co films had junctions with a 30° misorientation angle at the grain boundary of the bicrystal substrate. The superconducting properties of the films were onset $T_c$ of 21.5 K and $J_c$ of $2.0 \times 10^6$ A/cm$^2$ at 4 K, which are comparable to those of a previous report [19].

## 3. Result and discussion

Figure 1 schematically depicts the structure of the dc-SQUID fabricated in a

T.Katase et al.

$BaFe_2As_2$:Co epitaxial film on a LSAT bicrystal substrate. The dc-SQUID structure was defined by photolithography and Ar ion milling. We employed 3-µm-wide junctions, which are narrower than those we used to observe the Josephson effect [19], because the dc-SQUID operation requires a rather low critical current ($I_c$) less than 200 µA. As confirmed by optical microscopy, a superconducting loop with a slit area of $18 \times 8$ µm$^2$ was located across the LSAT bicrystal boundaries.

The current–voltage (*I*–*V*) characteristics of the dc-SQUIDs were examined using the four-probe method. The electrical contacts with the patterned film were made using In metal pads and Au wires, as shown in Fig. 1. A magnetic flux was applied perpendicular to the film surface with a solenoid coil for voltage–flux (*V*–*Φ*) measurements. The *V*–*Φ* characteristics were recorded using an oscilloscope by applying a constant bias current to the dc-SQUIDs, while the noise levels of the dc-SQUIDs were measured using a flux locked loop (FLL) circuit with an equivalent input voltage noise of 0.4 nV/Hz$^{1/2}$. To avoid interference from an environmental flux, the dc-SQUIDs were placed inside a µ-metal magnetic shield cup.

Figure 2(a) shows the *I*–*V* characteristics of the dc-SQUID measured at 13 K. The $I_c$ of the dc-SQUID was 350 µA, which corresponds to an average $J_c$ for each junction of 19 kA/cm$^2$. The *I*–*V* curve did not exhibit hysteresis. The inset, which is the theoretical fit to the Ambegaokar-Halperin (AH) model that considers the effect of thermal fluctuation [22], demonstrates that the AH model explains the experimental *I*–*V* curve. This agreement confirms that the junctions in the dc-SQUID exhibit a resistively-shunted junction behavior with a minimal excess current. Moreover, the fit provided the ratio of the junction coupling energy to the thermal energy $\gamma = \hbar I_c / e k_B T^* \approx 160$, where $T^*$ is the effective noise temperature (in this case $T^* \approx 104$

K), $k_B$ the Boltzman constant, $e$ the elementary electric charge, and $\hbar$ the Planck constant. We believe that the high value of $T^*$ was due to noise coupled in from room temperature sources. Considering the estimated junction resistance $\frac{R_N}{2}$ of 0.021 Ω from the theoritical I–V curve, we estimated the $I_cR_N$ value at this temperature was 7.35 μV.

Figure 2(b) shows the temperature dependence of $I_c$. $I_c$ monotonically increased as the tempeature decreased. The inset shows the $I_cR_N$ product, which was 48 μV at 4 K but rapidly decreased as the temperature increased. The temperature dependence and the magnitude of the parameters are similar to those of the single Josephson junctions reported in ref. 19 and agree with a superconductor–normal metal–superconductor (SNS) type junction behavior [23].

Figure 3 shows the V–Φ characteristics of the dc-SQUID measured at 14 K. The bias current of the dc-SQUID was adjusted to give a maximum voltage modulation depth. Periodic modulation was clearly observed. The shape of the V–Φ curve slightly depended on the direction of the applied magnetic field, and deviated from the ideal sinusoidal curve due to the asymmetric $I_c$ magnetic field dependence of the dc-SQUID. These behaviors have also been observed in single Josephson junctions [19]. The origin can be attributed to the non-uniform $I_c$ distribution between the two junctions and/or at the bicrystal grain boundaries.

The modulation voltage $\Delta V$ (defined as $V(\Phi = \frac{\Phi_0}{2}) - V(\Phi = 0)$) was 1.4 μV. The estimated geometrical SQUID inductance $L$ was 13 pH [24] using a magnetic penetration depth $\lambda$ of ~300 nm [25]. Furthermore, $\Delta V$ was calculated from the theoretical expression $\Delta V = \frac{4}{\pi} \frac{I_c R_N}{1+\beta} \left[ 1 - 3.57 \frac{\sqrt{k_B T L}}{\Phi_0} \right]$ in the presence of thermal noise, where $\beta = \frac{2 L I_c}{\Phi_0}$ is the inductance parameter [26], as 1.5 µV, which agrees well with the experimentally-obtained value. As the temperature decreased from 15 to 13 K, $\Delta V$ increased from 1.2 to 1.6 µV.

Figure 4 shows the flux noise $S_\Phi^{1/2}$ spectrum as a function of frequency $f$ for the dc-SQUID measured at 14 K using the dc bias mode of the FLL circuit. In the white noise region of $f > 20$ Hz where $S_\Phi^{1/2}$ is virtually constant against $f$, a flux noise level of $1.2 \times 10^{-4}$ $\Phi_0/\text{Hz}^{1/2}$ was obtained (solid line). On the other hand, the low frequency region of $f < 20$ Hz is explained well by a $1/f$ behavior (dotted line). The flux noise level at 1 Hz was ~$4.3 \times 10^{-4}$ $\Phi_0/\text{Hz}^{1/2}$. Large signals at 50 Hz and its harmonics are attributed to the power source coupling to the SQUID either magnetically or via the leads attached to it.

The obtained flux noise levels of the dc-SQUID in the $BaFe_2As_2$:Co epitaxial films were an order of magnitude higher than those of typical dc-SQUIDs using $YBa_2Cu_3O_{7-\delta}$ (YBCO) epitaxial films, which exhibit $\Delta V = 10$–20 µV at 77 K [27, 28]. The total flux noise density $S_\Phi$ of the dc-SQUID, which includes an equivalent input noise of 0.4 nV/Hz$^{1/2}$ in a FLL circuit, is given by $S_\Phi = S_{\Phi,\text{intrinsic}} + \frac{S_{V,\text{amp}}}{V_\Phi^2}$, where $S_{\Phi,\text{intrinsic}}$ is the intrinsic flux noise density of the dc-SQUID and $S_{V,\text{amp}}$ is the voltage noise density of the preamplifier. $V_\Phi = \frac{\partial V}{\partial \Phi}$ is the transfer coefficient and can be approximated as $\pi \Delta V$ if a sinusoidal modulation curve is assumed. Because $\Delta V$ obtained for the present dc-SQUID was as small as 1.4 µV, the influence of the preamplifier noise ($\left(\frac{S_{V,\text{amp}}}{V_\Phi^2}\right)^{1/2}$ = 9.1 × 10$^{-5}$ $\Phi_0$/Hz$^{1/2}$) must be considered. The estimated intrinsic flux noise level $S_{\Phi,\text{intrinsic}}^{1/2}$ was 7.8 × 10$^{-5}$ $\Phi_0$/Hz$^{1/2}$, but increased to 4.2 × 10$^{-4}$ $\Phi_0$/Hz$^{1/2}$ as the frequency decreased to 1 Hz.

We then focused on the origin of the intrinsic noise level. In the present dc-SQUIDs, which used BaFe$_2$As$_2$:Co epitaxial films, $I_c$ rapidly increased with decreasing temperature as shown in Fig. 2(b), but dc-SQUID operations using the FLL circuit required $I_c$ < 200 µA. Thus, the operation temperatures are limited to near $T_c$ of the dc-SQUID, and the operation is restricted to a small Δ$V$. The intrinsic flux noise density $S_\Phi$ of the dc-SQUID in the white noise region is given by $S_\Phi = \frac{16 k_B T R_N}{V^2}$ [29, 30]. Since $V$ scales with $R_N$, increasing $R_N$ will result in lower flux noise. Therefore, further narrowing the junction width so that a dc-SQUID device operates at lower temperatures should improve the flux noise level and Δ$V$. As noted in our previous report [19], the steep temperature dependence of $I_c$ and the rather low junction resistance (i.e., $R_N$) are due to the metallic nature of the normal-state BaFe$_2$As$_2$:Co. Thus, employing a superconductor–insulator–superconductor (SIS) junction structure should drastically improve dc-SQUID performances.

The present dc-SQUIDs displayed a 1/$f$ noise behavior at frequencies below 20 Hz, with a value of 4.2 × 10$^{-4}$ $\Phi_0$/Hz$^{1/2}$ at 1 Hz. This value is a factor of about 40 higher than that achieved in ref. [30], (1 × 10$^{-5}$ $\Phi_0$/Hz$^{1/2}$). At this point, we are unable to state whether the observed 1/$f$ noise arises from critical current noise or flux noise. In the future, we plan to perform bias reversal measurements to distinguish between these two sources of 1/$f$ noise [31].

## 4. Conclusions

In conclusion, dc-SQUIDs were successfully fabricated in $BaFe_2As_2$:Co epitaxial films grown on LSAT bicrystal substrates with 30° misorientation angles. Observing a clear periodic modulation in the $V$-$\Phi$ characteristics confirmed the dc-SQUID operation. The flux noise $S_\Phi^{1/2}$ spectra showed that the white noise level of the present dc-SQUID is an order of magnitude higher than typical dc-SQUIDs using YBCO epitaxial films, but this increase is primarily attributed to the small voltage modulation depth caused by the metallic nature of the normal-state $BaFe_2As_2$:Co. Employing a higher-resistivity junction structure should reduce the noise level.

**Acknowledgments**

This work was supported by the Japan Society for the Promotion of Science (JSPS), Japan, through "Funding Program for World-Leading Innovative R&D on Science and Technology (FIRST) Program".

**Figure captions**

Figure 1 Schematic dc-SQUID structure fabricated on LSAT bicrystal substrates with 30° misorientation angles. SQUID loop is located across the bicrystal grain boundary. In metal layers and Au wires were used for electrical contacts.

Figure 2 (a) Current–voltage ($I$–$V$) characteristics of the dc-SQUID measured at 13 K. Inset compares the theoretical fit (blue line) to the experimental curve (red dots). (b) Temperature ($T$) dependence of the critical current ($I_c$) for the SQUID (blue circles). Inset shows the temperature dependence of the $I_c R_N$ product (blue triangles).

Figure 3 Voltage–flux ($V$–$\Phi$) characteristics of the dc-SQUID measured at 14 K.

Figure 4 Frequency ($f$) dependence of the total flux noise $S_\Phi^{1/2}$, including an equivalent input noise in the FLL circuit, measured at 14 K for the dc-SQUID. $1/f$ noise behavior is observed at $f < 2$ Hz (the dotted line), whereas white noises are observed in the higher $f$ region (the solid line).

T.Katase et al.

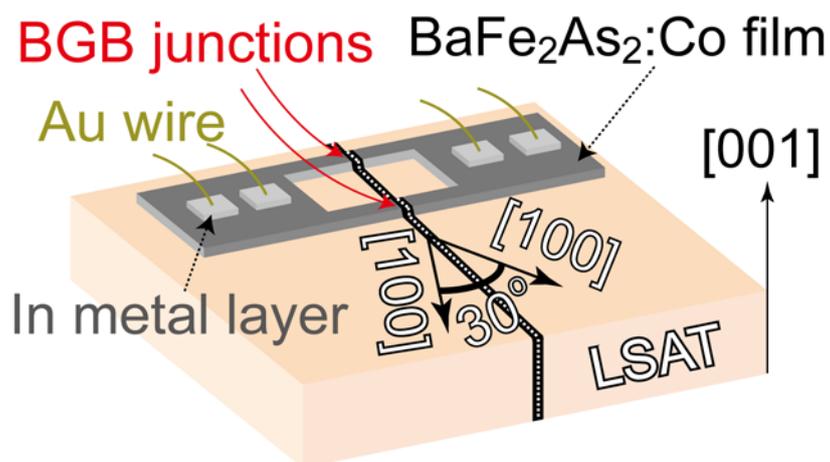

Figure 1.

T.Katase et al.

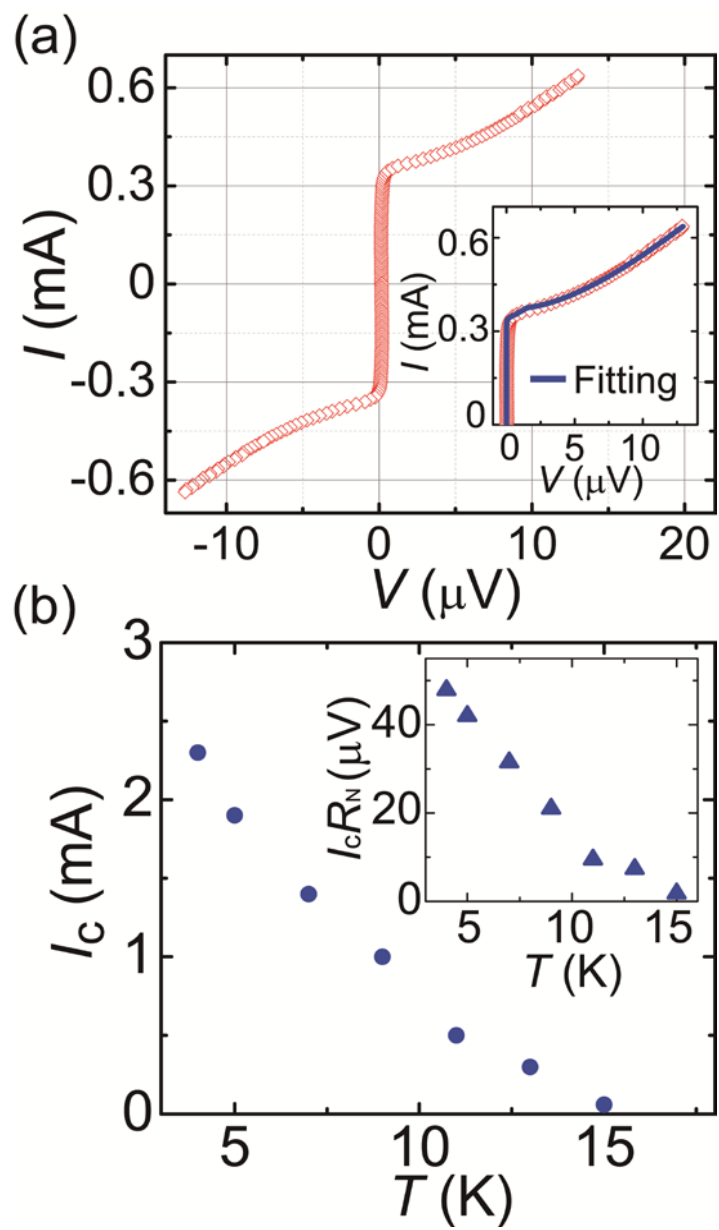

Figure 2.

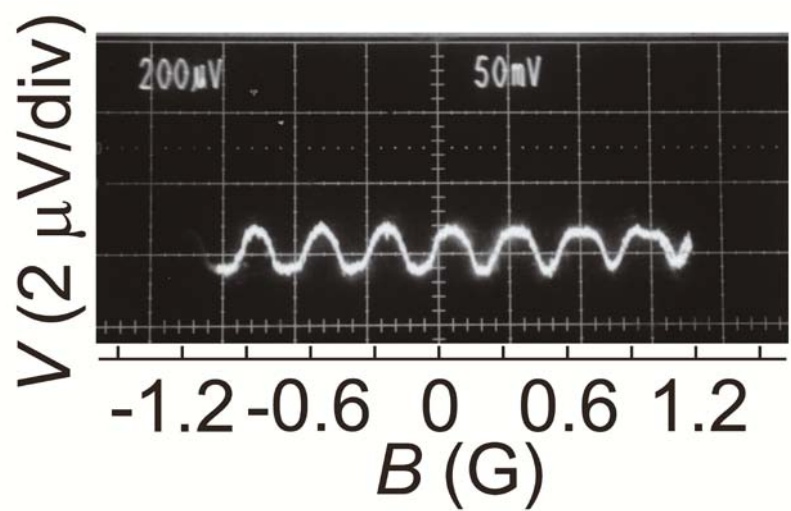

Figure 3.

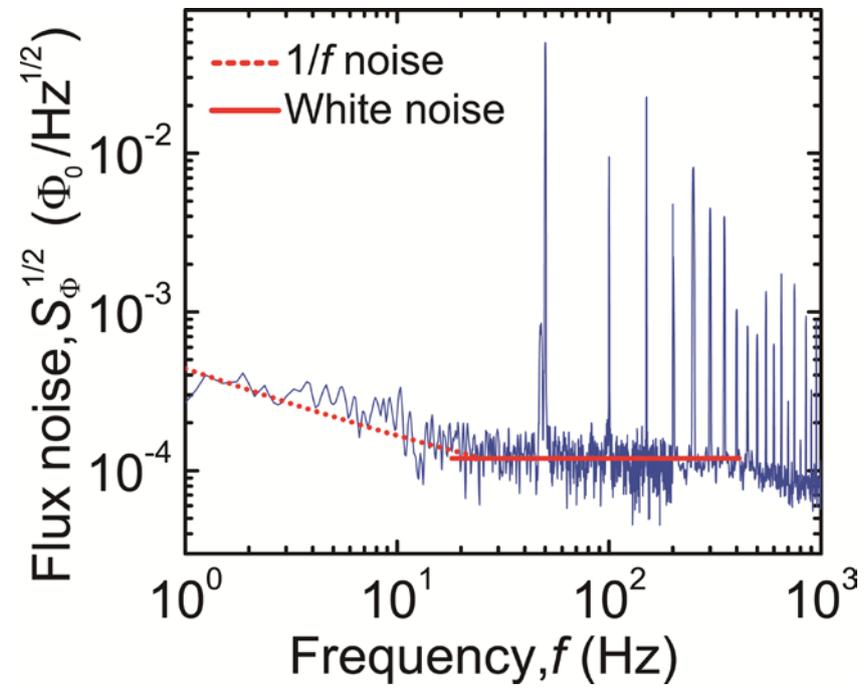

Figure 4.